\documentclass[12pt]{article}
\usepackage{graphicx} 
\usepackage{cite}
\input epsf.tex
\newcommand{\abs}[1]{\left | #1 \right |} 
\newcommand{\beq}{\begin{equation}}
\newcommand{\eeq}{\end{equation}}

\newcommand{\vev}[1]{ \left\langle {#1} \right\rangle }
\newcommand{\be}{\begin{equation}}
\newcommand{\ee}{\end{equation}}
\newcommand{\bea}{\begin{eqnarray}}
\newcommand{\eea}{\end{eqnarray}}
\begin{document}

\begin{titlepage}

\begin{center}

%\vspace{2cm}

{\hbox to\hsize {\hfill  UCB-PTH-04/19 }}
{\hbox to\hsize {\hfill  LBNL-49279 }}
{\hbox to\hsize {\hfill  SCIPP-2004/05 }}

\vspace{1cm}

{\Large \bf Localized Supersoft Supersymmetry Breaking}

\bigskip

{\bf Z. Chacko}$^{\bf a,b}$, {\bf Patrick J. Fox}$^{\bf c}$
and {\bf Hitoshi Murayama}$^{\bf a,b,d}$ \\

\medskip

$^{\bf a}${\small \it Department of Physics, University of California,
Berkeley, CA 94720, USA \\
\medskip
$^{\bf b}$ Theoretical Physics Group, Lawrence Berkeley National Laboratory,
\\  Berkeley, CA 94720, USA \\ 
\medskip
$^{\bf c}$ Santa Cruz Institute for Particle Physics,
     Santa Cruz CA 95064, USA \\
\medskip
$^{\bf d}$ School of Natural Sciences, Institute for Advanced Study,
Princeton, NJ 08540, USA\\
\medskip
%$^{\bf c}${\rm email}: zchacko@thsrv.lbl.gov \\
%$^{\bf d}${\rm email}: pjfox@scipp.ucsc.edu \\
%$^{\bf f}${\rm email}: murayama@hitoshi.berkeley.edu \\
}

\vspace{0.5cm}

{\bf Abstract}

\end{center}
\noindent

We consider supersymmetry breaking models in which the MSSM is extended to
include an additional chiral adjoint field for each gauge group with which the
the MSSM gauginos acquire Dirac masses. We investigate a framework in which
the Standard Model gauge fields propagate in the bulk of a warped extra
dimension while quarks and leptons are localized on the ultraviolet brane. The
adjoint fields are localized on the infrared brane, where supersymmetry is
broken in a hidden sector.  This setup naturally suppresses potentially large
flavor violating effects, while allowing perturbative gauge coupling
unification under SU(5) to be realized. The Standard Model superpartner masses
exhibit a supersoft spectrum. Since the soft scalar masses are generated at
very low scales of order the gaugino masses these models are significantly
less fine-tuned than other supersymmetric models. The LSP in this class of
models is the gravitino, while the NLSP is the stau.  We show that this theory
has an approximate R symmetry under which the gauginos are charged. This
symmetry allows several possibilities for experimentally distinguishing the
Dirac nature of the gauginos.

\end{titlepage} 
\renewcommand{\thepage}{\arabic{page}}
\setcounter{page}{1} 

\section{Introduction} 

While supersymmetry is the most attractive solution to the hierarchy problem
it nevertheless introduces several naturalness puzzles of its own.  Chief
among them is the `supersymmetric flavor problem' -- why do the squark masses
conserve flavor? Solutions to the supersymmetric flavor problem have come in
two forms -- either the squark masses have a renormalization group invariant
form that is nearly flavor diagonal at low
scales{\cite{Randall:1998uk,Giudice:1998xp,Jack:2000cd,Arkani-Hamed:2000xj}}
or the squark masses are finite and flavor diagonal at the scale at which they
are generated (for example,{\cite{dineetal,Kaplanet:1999ac,Chacko:1999mi}}).

Recently a new and attractive solution to this problem that falls into the
latter category has been proposed {\cite{Fox:2002bu}}. In this approach
the MSSM is extended to include a new chiral adjoint, often referred to as
an ``Extended Superpartner'' (ESP), for each SM gauge
group and an additional U(1$)'$ vector superfield under which the MSSM
fields are all singlets. The D component of this new U(1$)'$ is non-zero,
breaking supersymmetry. In such ``Gauge Extended Models'' (GEMs) each MSSM
gaugino acquires a Dirac mass with its corresponding adjoint through a
superpotential interaction involving this new U(1$)'$ vector superfield. The
scalar superpartners acquire their masses from loop diagrams that involve the
gauginos and the adjoints and which are dominated by momenta of order the
gaugino masses. The scalar masses are finite, loop suppressed with respect to
the gaugino masses and flavor diagonal at the matching scale. This pattern of
superpartner masses is called supersoft.  The fact that in this scenario the
gauginos are Dirac makes this an interesting
alternative to the conventional MSSM framework with Majorana gauginos.

In GEMs where the dominant source of supersymmetry breaking is
the non-zero D-term of a new vector superfield it is important to
understand if there are other effects which could cause the superpartner
masses to deviate from the supersoft form. Before addressing this question
let us first understand the origin of the supersoft contributions. The
superpotential operator that generates the gaugino masses has the form
\begin{equation} 
\label{eq:gauginomass} 
\int d^4 x \int d^2 \theta
\sqrt{2} \frac{W_{\alpha}' W^{\alpha} A}{M} + \mbox{h.c.} 
\end{equation}
where $W_{\alpha}'$ is the gauge field strength of the hidden sector
vector superfield whose D-component VEV $\vev{D'}$ breaks supersymmetry,
$W_{\alpha}$ is the gauge field strength of a Standard Model vector
superfield and $A$ is its corresponding ESP. This
leads to a Dirac gaugino mass $M_D \approx \; \vev{D'}/M$. The same
interaction results in supersoft scalar masses of order 
\begin{equation}
m_Q^2 \approx \frac{g^2}{16 \pi^2} M_D^2 
\end{equation} 
at one loop order.

Now if $M$ is of order the Planck scale, $M_{Pl}$, then realistic
phenomenology requires that $\sqrt{\vev{D'}} \approx 10^{11}$ GeV. Since
the natural size of any Fayet-Iliopoulos term is
expected to be of order $M_{Pl} \approx 10^{18}$
GeV~\cite{Fischler:1981zk} the more natural way to generate a non-zero 
VEV, $\vev{D'}$, of the
right size is to dynamically generate VEVs for a pair of fields $P$ and
$\bar{P}$ that are charged under the new U(1$)'$. If these VEVs $\vev{P}$
and $\vev{\bar{P}}$ (and their difference) are of order $10^{11}$ GeV then
$\vev{D'}$ naturally has the right size. This however immediately leads to
a problem - operators of the form 
\begin{equation}
\label{eq:flavorviolation} 
\int d^4x \int d^4 \theta \; \frac{P^{\dagger}
e^{V'} P \; Q^{\dagger} Q}{M^2} 
\end{equation} 
which are allowed by all
the symmetries of the theory give potentially flavor violating
contributions to the masses of the MSSM squarks $\tilde{Q}$ which are
larger than the supersoft contributions.

Can such flavor violating effects be avoided?  One possibility is that the
operator of Eqn.~(\ref{eq:gauginomass}) which gives rise to gaugino masses
only arises at loop level after integrating out heavy fields at some
intermediate scale $M$. However, as was shown in {\cite{Fox:2002bu}}, the
operator 
\begin{equation} 
\int d^2 \theta \frac{{W^{\alpha}}' W_{\alpha}' A^2}{M^2} +\mbox{h.c.}
\end{equation} 
is then typically generated at the same loop
order.  Although this operator does not give divergent contributions to
the soft scalar masses it is nevertheless problematic.  It
gives a negative contribution to the mass squared of either the scalar or
pseudoscalar component of the adjoint superfield that is larger (by a loop
factor) than the Dirac gaugino mass squared. If there is no other source 
of
supersymmetry breaking for $A$ a supersymmetric mass term for the adjoint
superfield 
\begin{equation} 
\int d^2 \theta M_A A^2 
\end{equation} 
must be
added to the theory to avoid charge and color breaking vacua. Although
this leads an acceptable superparticle spectrum, since $M_A$ is
significantly larger than $M_D$ the Dirac character of the gauginos is
lost, and the spectrum of the model tends toward that of intermediate
scale gaugino mediation {\cite{Csaki:2001em}},{\cite{Cheng:2001an}}.

From this discussion it is clear that the simplest models 
of Dirac gauginos based on the
scenario outlined in {\cite{Fox:2002bu}} do not solve the supersymmetric
flavor problem. What can be done to remedy the situation? The approach we will
take is to consider a five dimensional space where the MSSM matter fields and
the fields $P$ and $\bar{P}$ are localized on different 3-branes. Then the
operator of Eqn.~(\ref{eq:flavorviolation}) is forbidden by locality
\cite{Randall:1998uk}. Within such a framework there remain two distinct
possibilities; one is that the MSSM gauge fields propagate in the bulk of the
extra dimension while the other is that they too are localized on a brane with
the MSSM matter fields while the U(1$)'$ propagates in the bulk.  The
phenomenological implications of these two possibilities are NOT in fact the
same. In what follows we concentrate on the case where the MSSM gauge fields
are in the bulk of the space as this offers some significant advantages over
the other case, specifically in regard to electroweak symmetry breaking and 
coupling constant unification. We now explain these advantages.

%In models with a supersoft superpartner spectrum, obtaining electroweak
%symmetry breaking is typically not simple. 

%The reason is that for the case of MSSM gauge
%fields in the bulk one can write the brane localized operator

%\begin{equation} 
%\label{eq:hardbreaking} 
%\int d^5 x \sqrt{-G} \; \delta \left(y - \pi R \right) \int d^4 \theta
%\frac{W^{\alpha} D^2 W_{\alpha} P^{\dagger} e^{V'} P}{M_5^4} 
%\end{equation} 
%The effect of this and similar
%hard supersymmetry breaking operators is to alter the relative couplings
%of gauge bosons and gauginos to matter.  Related operators and their
%effects have been discussed in detail in~\cite{Kaplan:2000jz}. They cannot
%be forbidden by any symmetry, and generate flavor diagonal scalar masses
%at one loop of a size comparable to the supersoft contributions. These
%contributions are finite and cutoff at the compactification scale.

In models with a supersoft superpartner spectrum obtaining electroweak
symmetry breaking is typically not simple. The reason is that in the limit of
Dirac gaugino masses (which corresponds to $M_A \rightarrow 0$ above), the
D-terms which give rise to the Higgs quartic potential vanish, and therefore
the tree level Higgs mass vanishes. Additional contributions to the Higgs mass
from stop loops can give rise to realistic phenomenology even for $M_A <<
M_D$, but only if the stops are very heavy, which tends to push up the entire
spectrum of masses leading to fine-tuning.  In our extra dimensional 
scenario however, if the gauge
fields are in the bulk the scalar adjoints can be localized to the brane where
supersymmetry is broken. This allows them to directly acquire tree level
supersymmetry breaking masses from the operator
\begin{equation} 
\int d^5 x \sqrt{-G} \delta \left(y - \pi R \right)  \int d^4 \theta \;  
\frac{P^{\dagger} e^{V'} P \; A^{\dagger} A}{M_5^4} 
\label{PT}
\end{equation}
Now that the scalar adjoints are heavy the Higgs quartic potential takes
its familiar MSSM form, allowing for more natural electroweak symmetry
breaking. The low scale at which the supersoft scalar masses are generated
means that the negative correction to the Higgs soft masses from stop
loops does not get a large logarithmic enhancement.  Therefore the
fine-tuning required to obtain electroweak symmetry breaking is
significantly less than in, for example, the constrained MSSM.

Another advantage of the higher dimensional scenario with bulk gauge
fields relates to gauge coupling unification. Gauge coupling
unification must proceed differently in this class of models even in
four dimensions because of the additional adjoints. For unification
under SU(5) the SU(3), SU(2) and U(1) adjoints $A$ must appear at low
energies as part of a complete adjoint of SU(5). However the presence
of all these additional fields at low scales makes the gauge couplings
sufficiently large before unification is achieved that the
predictivity of the model is destroyed.  (This difficulty can be
avoided if the unifying group is $SU(3)^3$, in which case the extra
matter content is small enough to allow perturbative unification
{\cite{Fox:2002bu}}.) The higher dimensional scenario, however, admits
perturbative unification under SU(5)
if the fifth dimension is warped~\cite{Randall:1999ee} and the extra
adjoints localized to the infrared brane (along with the U(1$)'$ gauge
field) while the gauge symmetry is broken on the ultraviolet
brane. This is because in such a scenario the adjoint fields do not
contribute to running above the infrared cutoff. Proton decay can be
avoided by having the chiral matter of the MSSM on the ultraviolet
brane, which implies that supersymmetry must be broken on the infrared
brane in order to avoid the flavor violating operators of
Eqn.~({\ref{eq:flavorviolation}}), and to allow the operator of
Eqn.~({\ref{PT}}).  We see that the three requirements of unification
under SU(5), a flavor diagonal sparticle spectrum and the absence of
rapid proton decay together tightly constrain the locations of the
various fields in the higher dimensional space.

What are the distinguishing characteristics of this class of models?  
Since the scale at which supersymmetry is broken is warped down the LSP in
this class of models is the gravitino while the NLSP is usually the
(right-handed) stau. A very interesting feature of this class of models is
that they possesses an approximate R symmetry under which the gaugino and
antigaugino have opposite charges.  An important consequence of this
symmetry is that even though the gauginos are relatively heavy it may
nevertheless be straightforward to experimentally distinguish them from
Majorana gauginos provided the staus are stable on collider timescales.
The reason is that decays of pair produced superparticles nearly always
result in oppositely charged staus in the final state if the gauginos are
Dirac whereas the staus often have the same charge if the gauginos are
Majorana. Even if the staus are not stable on collider time scales it may
still be possible to distinguish between Dirac and Majorana gauginos by
accurately measuring the charges and energies of the muon and two taus
produced in smuon decays. These measurements enable the charge of the
decaying smuon to be related to that of the resulting (short-lived) stau -
these almost always have the same charge if the gauginos are Dirac.
In the sections which follow we explain our model in greater
detail.

\section{The Framework}

In this section we establish the framework we will be working in and the
notation we are using. We
consider a five dimensional setup.  We employ a coordinate system $x^M$
where $M$ runs from 0 to 3 and 5.  The fifth dimension $x^5 = y$ is
compactified on the interval $0 \le y \le \pi R$, which can be thought as
arising from the orbifold $S^1/Z_2$. There are 3-branes at the orbifold
fixed points $y =0$ and $y = \pi R$ on which fields are localized.  The
metric is given by the line element 
\begin{equation} 
\label{lineelement}
ds^2 = G_{MN} dx^M dx^N = e^{-2 \sigma} \eta_{\mu \nu} dx^{\mu}dx^{\nu} +
dy^2~. 
\end{equation} 
Here the $x^{\mu}$, where $\mu$ runs from 0 to 3,
parametrize our usual four spacetime dimensions, $\eta_{\mu \nu} =
{\rm{diag}}(-1,1,1,1)$, and $\sigma = k \abs{y}$, where $k$ is the AdS
curvature and is related to the four dimensional Planck scale $M_{4}$ and
the five dimensional Planck scale $M_{5}$ by 
\begin{equation}
\label{4DPlanckscale} 
M_4^2 = \frac{M_5^3}{2k}\left(1 - e^{-2 k \pi R}
\right) \simeq \frac{M_5^3}{2k}~, 
\end{equation} 
where the second equality holds when the warping is significant.

For what follows we will need to understand the couplings of supersymmetric
bulk vector multiplets and bulk hypermultiplets~\cite{Gherghetta:2000qt}. An on
shell vector multiplet in five dimensional ${\cal{N}} = 1$ supergravity
consists of a gauge field $A_M$, a pair of symplectic Majorana spinors
$\lambda^i$, with $i=1,2$, and a real scalar $\Sigma$ which transforms in the
adjoint representation.  The bosonic part of the higher dimensional gauge
field action takes the form\footnote{The minus sign in front of the scalar
  kinetic term is due to our metric signature convention,
  Eqn.~(\ref{lineelement}).}

\begin{equation} 
S_{b} = -\int d^4x \int dy \sqrt{-G}\frac{1}{g_5^2} \left[\frac{1}{4} F_{MN}
F^{MN} + \frac{1}{2} D_M \Sigma \; D^M \Sigma + \left( \sigma'' -2 k^2
\right) \Sigma^2 \right]~, 
\end{equation}
where $\sigma'' = 2k \left[\delta(y)-\delta(y-\pi R) \right]$. The
fermionic part takes the form

\begin{equation} 
\label{gauginoaction}
S_{f} = -\int d^4 x\int dy\sqrt{-G}
\frac{i}{2g_{5}^{2}} \left[\bar{\lambda}^i \Gamma^M D_M \lambda^i +
\frac{1}{2} \sigma' \bar{\lambda}^i \left( \sigma^{3} \right)^{ij}
\lambda^j \right]~, 
\end{equation}
where $D_{M}$ is a covariant derivative with respect to both general coordinate
and gauge transformations.  The vielbein factors necessary to write the spinor
action in curved space are implicit in Eqn.~(\ref{gauginoaction}).

We demand that $A_{\mu}$ and $\lambda^1_L$ are even while $\Sigma$ and
$\lambda^2_L$ are odd.  Then the even fields each have a massless mode
with the following $y$ dependence:
\begin{eqnarray}
A_{\mu}\left(x,y\right) &=& \frac{1}{\sqrt{\pi R}} A_{\mu}^{\left(0\right)}
\left(x\right) + \cdots \\
\label{gauginozeromode}
\lambda^1_L\left(x,y\right) &=& \frac{e^{3\sigma/2}}  {\sqrt{\pi R}}
\lambda^{1 \; \left(0 \right)}_L\left(x\right) + \cdots
\end{eqnarray}

The lightest KK masses are of order $M_c = k\,e^{-k \pi R}$, which we will
call the compactification scale. We are interested in compactification
scales larger than the scale of supersymmetry breaking but much
smaller than the unification scale $\approx 10^{16}$GeV.

A bulk hypermultiplet $H$ consists of two complex scalars $\phi^i$ and a
Dirac fermion $\psi$. The bulk action has the form 
\begin{equation}
S_H =-\int d^4 x\int dy \sqrt{-G} |\partial_M \phi^i|^2 + i \bar{\psi}\Gamma^M
D_M \psi + m_{\phi,i}^2 |\phi^i|^2 + i m_{\psi} \bar{\psi} \psi 
\end{equation} 
where the five dimensional masses of the scalars and the fermion are
constrained to satisfy 
\begin{eqnarray} m_{\phi,i}^2 &=& \left(c^2 \pm c - \frac{15}{4}
\right)k^2 + \left(\frac{3}{2} \mp c \right) \sigma'' \\ m_{\psi} &=& c
\sigma' 
\end{eqnarray}
Here $c$ is a dimensionless number that can be
chosen arbitrarily.  We demand that $\phi^1$ and $\psi_L$ are even, while
$\phi^2$ and $\psi_R$ are odd. Then each even field has a massless mode
with a profile given by
 \begin{eqnarray}\label{eqn:hyperprofiles}
 \phi^1 \left(x, y \right) &=&
{e^{\left(3/2 - c\right)\sigma}}\phi\left(x \right) \\ \psi_L \left(x, y
\right) &=& {e^{\left(2 - c\right)\sigma}}\psi\left(x \right)  
\end{eqnarray}
For $c = 1/2$, the kinetic terms are independent of the
extra coordinate, just as the zero modes of the bulk vector multiplet.

\section{The Model}

In this section we describe our model in detail. We are interested in a
scenario where the Standard Model quark, lepton and Higgs fields are localized
on the brane at $y =0$ where the warp factor is large and the local scale is
$\sim M_{GUT}$ while supersymmetry is broken on the brane at $y = \pi R$ where
the warp factor is small and the local scale is $\sim M_{IR}$.  The MSSM gauge
fields live in the bulk and couple to both the matter fields localized on the
ultraviolet brane, and to the supersymmetry breaking sector which is localized
on the infrared brane{\footnote{For earlier work on supersymmetry breaking in
    warped extra dimensions see, for example,
    \cite{Gherghetta:2000qt,Gherghetta1,Goldberger:2002pc,Choi:2003fk,Chacko:2003tf,Choi:2003di,Nomura:2003qb,Nomura:2004is}}.
  We assume that the grand unifying symmetry, which is SU(5), is broken on the
  UV brane by the Higgs mechanism. The ESPs $A_i$ are also localized on the
  infrared brane.  Since the SU(5) symmetry is unbroken there these form a
  complete SU(5) multiplet. The interaction which gives the
gauginos a mass takes the familiar supersoft form %
\begin{equation} 
\int d^5 x \sqrt{-G} \; \delta \left(y - \pi R \right)
\int d^2 \theta \sqrt{2} \frac{W_{\alpha}' W_i^{\alpha} A_i}{M_5}
\end{equation}
where $W_i$ is the field strength, for the $i^{th}$ gauge group, formed from
the even components of the bulk gauge supermultiplet. $W'$ is the field
strength of the U(1$)'$ gauge field which is localized on the IR brane and
whose D component $\vev{D'}$ breaks supersymmetry. This leads to a gaugino
mass 
\begin{equation}
M_{D, i}(\mu) = \frac{\vev{D'}}{M_5} e^{-k \pi
  R}\left[\frac{\alpha_i(\mu)}{\alpha_i(M_{IR})}\right]^{\frac{b_i-2c_i}{2b_i}}
  \left[\alpha_i(M_{IR})\right]^{1/2},
\end{equation}
where $b_i=-(3N_c-N_f)$ is the coefficient of the beta function for the gauge
coupling $g_i$ and $c_i$ is the quadratic Casimir of the gauge representation,
$c_i=(N^2-1)/2N$ for a fundamental representation.  In the limit that the the
infrared scale is very close to the weak scale each gaugino mass is
proportional to its corresponding gauge
coupling{\footnote{Remarkably, these models share this feature with warped
    five-dimensional models with strong F-term supersymmetry breaking
    localized on the infra-red brane\cite{Goldberger:2002pc,Chacko:2003tf,Nomura:2003qb,Nomura:2004is}. In the limit of
    strong supersymmetry breaking the gauginos in these models are also
    pseudo-Dirac, with masses proportional to the corresponding gauge coupling
    \cite{Chacko:2003tf}. However the scalar masses are in general not the
    same in the two models even in this limit because of
    the second term in the logarithm in Eqn.(\ref{eq:ssscalar}).}}.  Note
that here $\vev{D'}$ is defined as a scale in the 5D theory.  In the effective
4D theory this scale is warped down, Eqn.(\ref{eqn:appendixdiracmass}).

The MSSM scalar superpartners acquire a supersoft contribution to their soft
masses that arises at loop level from the gaugino mass $M_D$. This
contribution is finite and given by~\cite{Fox:2002bu},
\begin{equation}\label{eq:ssscalar}
 {m^2}_{i,SS}(\mu) = \frac{c_i \alpha_i M_{D,i}^2(\mu)}{\pi} \log
\left(4
%M_{D,i}^2(\mu)
+\frac{m_{\tilde{A}}^2}{M_{D,i}^2(\mu)}\right)
\end{equation}
where $M_{D,i}$ is the
mass of the gaugino of the $i$th gauge group and $m_{\tilde{A}}^2$ is the
mass$^2$ of the scalar adjoint.

We assume for the purposes of this model that the non-zero VEV $\vev{D'}$ is
generated dynamically as the difference in the VEVs of a pair of fields $P$
and $\bar{P}$ which have opposite charges under the U($1)'$. This is more
natural than employing a Fayet-Iliopoulos term. If there are no unnaturally
small parameters in this sector we expect $\vev{P} \; \approx \; \vev{\bar{P}}
\; \approx \; \sqrt{\vev{D'}}$. The scalar component of the adjoint field A
can then directly pick up a soft mass ${m_{\tilde{A}}}^2$ from the interaction
\begin{equation} 
\int d^5 x \sqrt{-G} \; \delta \left(y - \pi R \right)
\int d^4 \theta \frac{P^{\dagger} e^{V'} P A^{\dagger} A}{M_5^2}
\label{Asoftmass}
\end{equation}
Since this arises directly at tree level and has no large volume suppression
we expect that the scalar components of the adjoint will be relatively heavy.
This soft mass for the scalar adjoints feeds back into the soft masses of
the MSSM scalars at two loops through gauge interactions \cite{Poppitz:1997xw}.
\begin{equation}
\label{bad}
\Delta{m^2} \approx - \left(\frac{g^2}{16 \pi^2}\right)^2 {m_{\tilde{A}}}^2
{\rm log}\left(\frac{k e^{-k \pi R}}{m_{\tilde{A}}}\right)
\end{equation} 
This term can give sizable corrections to the soft scalar masses that are
comparable to the supersoft contribution. A `Naive Dimensional Analysis'
(NDA) estimate (see the appendix) suggests that the size of these
corrections can naturally be limited to about 20\% of the supersoft
contribution.  If however the infrared scale is high, the logarithm
becomes important and mild finetuning may be needed.

Another potential contribution to the soft scalar masses arises from hard
supersymmetry breaking operators localized on the infrared brane. These have
the form
\begin{equation}
\int d^5 x \sqrt{-G} \; \delta \left(y - \pi R \right)
\int d^4 \theta \frac{W^{\alpha} D^2 W_{\alpha} \; P^{\dagger} e^{V'}
P}{M_5^4}
\label{hardbreaking}
\end{equation}
The effect of this and analogous operators is to alter the relative couplings
of gauge bosons and gauginos to matter.  In the context of higher dimensional
models similar operators were considered in \cite{Kaplan:2000jz}.  They cannot
be forbidden by any symmetry, and generate flavor diagonal scalar masses at
one loop.  However an NDA estimate suggests that these corrections are small
compared to the supersoft contribution, as shown in the appendix.

Yet another potential contribution to the soft scalar masses arises
from the kinetic mixing between U(1)$^\prime$ and U(1)$_Y$ gauge fields.
If it is unsuppressed,
\begin{equation}
\int d^5 x \sqrt{-G} \; \delta \left(y - \pi R \right)
\int d^2 \theta\  W'_{\alpha} W_Y^{\alpha},
\label{kineticmixing}
\end{equation}
will induce a large $D_Y \approx D$ for $U(1)_Y$ and contributions to
scalar masses $\Delta m_i^2 = g^{\prime 2} Y_i D_Y$.  It overcomes the
one-loop supersoft contributions of Eq.~(\ref{eq:ssscalar}), and hence
some of the scalar masses become negative.  Our model avoids this
problem with the SU(5) invariance on the IR brane that does not allow
for this kinetic mixing.  Because U(1)$^\prime$ is localized on the IR
brane and there is no possible counter term, there is no
logarithmically divergent radiative effects either.  We conclude the
model is immune to this potential disaster.

Since the supersymmetry breaking scale is relatively low the LSP in these
models will be the gravitino. The NLSP will be the right handed stau,
which we expect to be somewhat lighter than the other right handed sleptons.
The large soft mass of the scalar adjoints implies that the Higgs quartic
terms are not suppressed as in minimal supersoft models. The relevant part 
of the scalar Lagrangian is
\begin{equation} 
L_S = -M_D ( \tilde{A} + \tilde{A^*} ) D - D
\left(\Sigma_i \; g \; q^*_i T q_i \right)
      -\frac{1}{2} D^2 - {m_{\tilde{A}}}^2 \left( \tilde{A} \tilde{A^*}\right)
\end{equation} 
From this it is straightforward to infer that the Higgs quartic term is
the same as in the MSSM upto corrections of order $M_D^2/m_{\tilde{A}}^2$.
Pushing $m_{\tilde{A}}$ high allows EWSB to go through as in the MSSM but
for large infrared scale this reintroduces the problem of large negative
contributions to the soft scalar massed squared, Eqn.(\ref{bad}). When
minimising the Higgs potential in subsequent sections we take into account
the small correction to the quartic coming from the triplet and singlet
ESP scalars.  For the purposes of this paper we treat $\mu$ and $B\mu$ as
free input parameters and do not specify a dynamical origin for them. It
may be possible to generate suitable values for these parameters from the
VEV of a singlet which lives in the bulk and communicates directly with
the supersymmetry breaking sector. However we do not pursue this
possibility further here.

Perturbative gauge coupling unification arises naturally in this model, even
under SU(5). The SU(3), SU(2) and U(1) ESPs are now extended to a complete
adjoint of SU(5). The GUT group, which for concreteness we take to be
SU(5), is assumed to be broken on the ultraviolet brane. As has recently been
established, gauge couplings run logarithmically in AdS
spaces~\cite{Pomarol:2000hp,Randall1,Agashe1,Choi1,Contino:2002kc,Goldberger1,Falkowski:2002cm,Randall:2002qr}.
Since the adjoint field is localized on the infrared brane its contribution to
running is SU(5) symmetric and cutoff at the compactification scale, while the
brane-localized gauge couplings on the ultraviolet brane which do not respect
SU(5) unify at the GUT scale.

An additional interaction is required to give a mass to the fermions in
the adjoint of SU(5) that are not adjoints
of SU(3), SU(2) or U(1)~\footnote{The ESPs in an adjoint of $SU(5)$ that are
not adjoints under the Standard Model are referred to as ``bachelors''.  The
spin-$1/2$ components of the bachelors are sometimes referred to as
``spinsters''~\cite{Nelsontasi}.}.  The scalar adjoints all acquire a mass
from the supersymmetry breaking interaction above while the fermions that are
adjoints under SU(3), SU(2) and U(1) acquire Dirac masses with the
gauginos. One possibility is to write a mass term 
\begin{equation} 
\int d^5 x \sqrt{-G} \; \delta \left(y - \pi R \right) \int d^2 \theta M_A A^2 
\end{equation} 
on the infrared brane. Since such a term is necessarily SU(5) symmetric this
will result in a mass for the SU(3), SU(2) and U(1) adjoint fermions as well
with the result that the gauginos are no longer purely Dirac. However provided
$M_A$ is sufficiently smaller than $M_D$ the gauginos will still be
pseudo-Dirac. As with $\mu$ and $B \mu$, for the purposes of this paper we
treat $M_A$ as a free parameter and do not specify a dynamical origin for it.

In the absence of the $\mu$, $B\mu$ and $M_A^2$ terms in the Lagrangian the
theory has an exact continuous R-symmetry under which all the chiral
superfields of the MSSM have charge 1 and the adjoint superfield $A$ and Higgs
fields have charge 0.  This approximate symmetry has several important
consequences. Since any A-term for the MSSM fields breaks this symmetry it can
only be generated by loops involving these couplings.

This approximate R symmetry also leads to very interesting experimental
signals for the Dirac gauginos in this class of models. Since the gauginos
are relatively heavy it might have been expected that it would be
extremely difficult to distinguish them from Majorana gauginos. However
this is not the case if the NLSP, which is the stau, is long lived.
Consider the decay of a pair of superparticles that have been pair
produced in a collider (Figures \ref{fig:lepton1} and \ref{fig:hadron}).
Each of these will eventually decay into a stau-tau pair primarily through
an off-shell bino. However since the bino and anti-bino have opposite
R-charges each of the two staus that is produced will always have opposite
sign charge relative to the other as will to its associated tau.  This is
in sharp contrast to the case where the gauginos are Majorana. Here, since
a gaugino Majorana mass term breaks the R symmetry, the two staus could
have the same charge. The staus are unstable and each will decay into a
tau and a gravitino. The stau lifetime depends on the supersymmetry
breaking scale $\sqrt{D'}$.  If the staus are stable on collider
time-scales their charges can be determined. This seems a very
promising approach to distinguish Dirac gauginos from Majorana gauginos in
this class of models.

\begin{figure}[t]
\centerline{\epsfxsize=3in \epsfbox{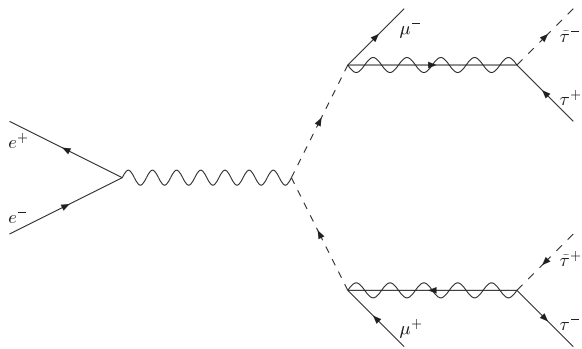}}
\caption{Signal of Dirac gauginos at a leptonic collider.}
\label{fig:lepton1}
\end{figure}

\begin{figure}[t]
\centerline{\epsfxsize=3in \epsfbox{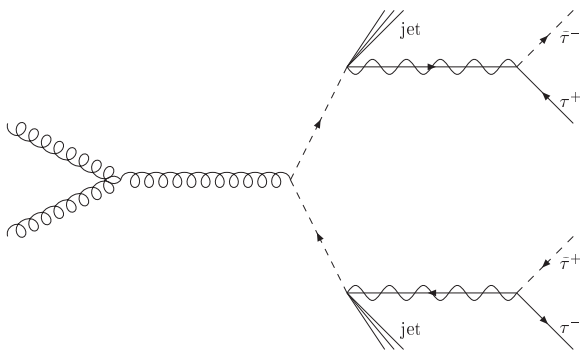}}
\caption{Signal of Dirac gauginos at a hadronic collider.}
\label{fig:hadron}
\end{figure}

Since in our model the R-symmetry is not exact the gauginos are therefore
only pseudo-Dirac, the staus will sometimes have the same charge. The
frequency with which this occurs depends on the ratio of the mass of the
SU(5)/SU(3)$\times$SU(2)$\times$U(1) adjoints to the gaugino mass. When
this ratio is very small the gauginos tend to be purely Dirac.

However in some models the supersymmetry breaking scale may be too low and
therefore the stau lifetime too short to allow the charges of the staus to
be determined. We now argue that it may nevertheless be possible to
distinguish Dirac gauginos from Majorana gauginos even in such a scenario.
A key observation is that in these models the mass splitting between the
right-handed sleptons is in general rather small. Consider then the decay
of a right-handed smuon that has been produced in a collider into a muon,
a tau and a stau through an (off-shell) bino.  The Dirac nature of the
gauginos implies that the stau has the same charge as the decaying smuon,
as does the muon.  The small mass splitting between the smuon and the stau
means that in the smuon rest frame the muon and the tau both have small
momenta. However in this frame the tau produced by the prompt decay of the
stau has a large momentum, and carries the same charge as the stau. If the
energies and charges of the smuon decay products can be accurately
determined in the laboratory frame, then by boosting back to the
(approximate) smuon rest frame it may be possible to distinguish between
the two taus. We can then determine if the smuon and the stau had the same
charge, and therefore (given enough events) if the gauginos are Majorana
or Dirac.

At lepton colliders Dirac and Majorana gauginos can also be distinguished by
using polarized beams. The approximate R-symmetry of these models implies that
the cross-sections for $e^{+} e^{-} \rightarrow$ 2 superparticles and also
$e^{-} e^{-} \rightarrow$ 2 superparticles are very small unless the incident
leptons have opposite helicities. Therefore a measurement of the cross-section
to two superparticles for different helicities of the incident leptons may be
sufficient to distinguish between the two cases.

For values of the supersymmetry breaking scale $\sqrt{\vev{D'}} \; e^{-k \pi
  R}$ greater than about 3$\times 10^{6}$ GeV the stau NLSPs are sufficiently
long lived that it may be possible to detect these particles in neutrino
telescopes {\cite{ABC}},{\cite{Bietal}}. Very high energy neutrinos
originating from astrophysical sources can collide with nuclei inside the
earth, producing a pair of superparticles which promptly decay to staus. These
staus can be seen in neutrino telescopes as a pair of parallel charged tracks
emerging from the earth about 100m apart.  Interestingly, the Dirac nature of
the gauginos in these models means that the expected number of events is
somewhat less than for Majorana gauginos even for the same superparticle
spectrum. The reason is that the R-symmetry restricts the helicities of the
incident (anti)neutrino and (anti)quark in these processes to be opposite,
while for the case of Majorana gauginos there is no such restriction. Given
enough events and some knowledge of the superpartner spectrum, this may be
another way to experimentally distinguish between Dirac and Majorana gauginos.
 
\subsection{Sample Spectra}

In the table below we give some sample spectra for this class of models.  For
simplicity we do not incorporate a dynamical origin for the $\mu$ and $B \mu$
terms but merely treat them as free input parameters.  The top Yukawa, $y_t$
is defined at the IR scale as is the gaugino mass parameter, $m_g\equiv
(D^\prime/M_5) e^{-k\pi R}$.  For simplicity the supersymmetric mass term 
for
the ESPs $M_A$ and the soft scalar mass for the ESPs $m_{\tilde{A}}$ are given
at 1 TeV.  For each set of inputs the superpartner mass spectrum is
calculated, the masses are given at 1 TeV.  The two sets of inputs, A and B,
correspond to an IR scale of 1000 TeV and 30 TeV respectively.

The deviation from the conventional MSSM Higgs potential is determined by
the magnitude of $M_D/m_{\tilde{A}}$, and has been included in the
spectrum. The size of the negative Poppitz-Trivedi contribution to scalar
mass squared relative to the supersoft contribution is set by the same
parameter, and also by $\log M_{IR}/m_{\tilde{A}}$.  For the points B the
Poppitz-Trivedi corrections are small and can be ignored.  For the points
A we do not include them but they can have a sizeable effect.  In both
cases we include the effect of the the ESPs on the Higgs potential, in
particular the suppression of the quartic term.

The final row in Table \ref{tab:spectra} measures the sensitivity of the Higgs
VEV to changes in $\mu$.  We search parameter space for regions where the
input parameters give correct EWSB with a Higgs VEV of 174 GeV.  At these
points we measure our sensitivity as,
\be
\kappa\equiv \frac{\mu}{v}\frac{\partial v}{\partial \mu}.
\ee
The sensitivity is clearly less than in typical high scale models of
supersymmetry breaking, for which $\kappa$ usually lies between 50
and 200. Therefore these supersoft models are significantly less 
fine-tuned than most supersymmetric models.

There are several general features of the spectrum we would like to
emphasise. The gauginos are pseudo-Dirac with small, $\mathcal{O}(M_A)$,
splittings. Colored particles are all heavier than 1.5 TeV and yet the
Higgs mass can still be light.  The stop loops that feed into the Higgs
mass are cutoff by the gluino mass rather than the usual GUT/Planck scale
since the mass for a colored scalar is only generated at the gluino mass.  
This finiteness property means we no longer have a log sensitivity to the
GUT scale, which is why these models are less fine-tuned than models where
the scalar masses are generated at high scales.

\begin{table}\label{tab:spectra}% [h] % htbp 
\centering 
\begin{tabular}{c|c|c|c|c|c} 
& & Point A & Point A$^\prime$ & Point B & Point B$^\prime$\\ 
\hline 
inputs:  & $m_{g}$ & 
$14\times 10^3$ & $18\times 10^3$ & $15\times 10^3$ & $21\times 10^3$\\
 & $m_{\tilde{A}}$ & $8.7 \times 10^3$ & $16\times 10^3$ & $16\times 10^3$ &
 $22\times 10^3$\\
 & $M_A$ & $300$ & $300$ & $300$ & $300$ \\
 & $y_t$ & 1 & 1 & 1.05 & 1.05\\
 \hline neutralinos: & $m_{\chi^0_1}$ & 426.1 & 556.0 & 323.8 & 446.6\\
 & $m_{\chi^0_2}$ & 426.3 & 556.1 & 324.0 & 446.7\\
 & $m_{\chi^0_3}$ & 1730 & 2240 & 1886 & 2611\\
 & $m_{\chi^0_4}$ & 2030 & 2540 & 2186 & 2911\\
 & $m_{\chi^0_5}$ & 2619 & 3371 & 2723 & 3748\\
 & $m_{\chi^0_6}$ & 2918 & 3671 & 3023 & 4048\\
 \hline charginos: & $m_{\chi^\pm_1}$ & 426.4 & 556.2 & 324.0 & 446.7\\
 & $m_{\chi^\pm_2}$ & 2617 & 3370 & 2722 & 3748 \\
 & $m_{\chi^\pm_3}$ & 2917 & 3670 & 3022 & 4048\\
 \hline Higgs: & $\tan\beta$ & 5.21 & 4.33 & 3.57 & 3.61\\
 & $m_{h^0}$ & 119.4 & 127.7 & 117.1 & 124.6\\
 & $m_{H^0}$ & 598.5 & 814.8 & 608.2 & 831.8\\
 & $m_A$ & 597.5 & 813.6 & 606.0 & 830.2\\
 & $m_{H^\pm}$ & 602.8 & 817.6 & 611.3 & 834.0\\
 & $\mu$ & 426.4 & 556.2 & 324.0 & 446.8 \\
 & $B$ & 155.0 & 260.7 & 294.6 & 397.1\\
 \hline sleptons: & $m_{\tilde{e}_R}$ & 199.6 & 273.8 & 242.5 & 328.4\\
 & $m_{\tilde{e}_L}$ & 407.1 & 569.5 & 490.4 & 667.4\\
 & $m_{\tilde{\nu}_L}$ & 401.8 & 564.4 & 484.8 & 663.3\\
 \hline stops: & $m_{\tilde{t}_1}$ & 1749 & 2393 & 1855& 2526\\
 & $m_{\tilde{t}_2}$ & 1708 & 2335 & 1860 & 2450 \\
\hline other squarks: & $m_{\tilde{u}_L}$ & 1740 & 2386 & 1847 & 2549 \\
 & $m_{\tilde{u}_R}$ & 1699 & 2339 & 1792 & 2445\\
 & $m_{\tilde{d}_L}$ & 1741 & 2388 & 1849 & 2520 \\
 & $m_{\tilde{d}_R}$ & 1696 & 2324 & 1787 & 2438\\
 \hline gluino: & $M_3$ & 6285 & 7998 & 5623 & 7634\\
\hline sensitivity: & $\kappa$ & 25 & 38 & 15 & 26
\end{tabular}
\caption{Sample points in parameter space, all masses are in GeV.  The A
  points have $M_{IR}=10^6$GeV and the B points have $M_{IR}=30\times 10^3$
  GeV.} 
\end{table}

\medskip

\section{Conclusions}

We have constructed models of supersoft supersymmetry breaking with Dirac
gauginos in which the breaking of supersymmetry is localized to a brane in
a higher dimensional space. These theories have three significant
advantages over the corresponding models in four dimensions -
\begin{itemize} 
{\item{natural suppression of potentially dangerous flavor
violating operators}} 
{\item{more natural electroweak symmetry breaking}}
{\item{the possibility of grand unification under SU(5)}}. 
\end{itemize}
In these theories the lightest supersymmetric particle is the gravitino,
while the next to lightest supersymmetric particle is the stau. The
spectra exhibit the characteristic supersoft relations among the
superparticle masses. In general the squarks and gauginos tend to be
fairly heavy, with masses of order a few TeV, while the sleptons are much
lighter. In spite of these large squark masses the degree of fine-tuning
required to obtain electroweak symmetry breaking is significantly better
than in most other supersymmetric models. The reason is that the scalar
masses are generated only at very low scales of order the gaugino masses,
and hence the negative correction to the Higgs soft mass squared from stop
loops does not have a large logarithmic enhancement. The large radiative
correction to the quartic term in the Higgs potential from the heavy stops
also partially compensates for the reduced contribution to the quartic
from the SU(2)$_{\rm L}$ and U(1)$_{\rm Y}$ D-terms that is a
characteristic of supersoft models.

Even if the gauginos are too heavy to be produced on shell an approximate
R-symmetry of these models implies that if the stau is stable on collider
time-scales the Dirac nature of the gauginos can be probed. The reason is
that decays of heavy superpartners almost always lead to opposite sign
staus in the final state if the gauginos are Dirac, while this is not true
for Majorana gauginos. Even if the stau is not stable on collider time
scales it may still be possible to distinguish between Dirac and Majorana
gauginos by accurately measuring the charges and energies of the muon and
two taus produced in smuon decays. These measurements enable the
charge of the decaying smuon to be related to that of the resulting
(short-lived) stau - these almost always have the same charge if the
gauginos are Dirac.

\bigskip

\noindent
{\bf Acknowledgements} \\

\noindent
ZC would like to thank Ian Hinchliffe for useful discussions. PF would
like to thank Neal Weiner and David E. Kaplan for useful discussions.
PF is supported in part by the U.S. Department of Energy. Z.C. and
H.M.  were supported in part by the Director, Office of Science,
Office of High Energy and Nuclear Physics, of the U.S. Department of
Energy under Contract DE-AC03-76SF00098 and DE-FG03-91ER-40676, and in
part by the National Science Foundation under grant PHY-00-98840.  The
work of H.M. was also supported by the Institute for Advanced Study,
funds for Natural Sciences.

\appendix
\section{Warped Naive Dimensional Analysis}

Consider a scenario where a theory becomes strongly coupled at some
fundamental scale $\Lambda$. Loop effects are now no longer suppressed
relative to tree level effects at this scale. The methods of `Naive
Dimensional Analysis' (NDA) \cite{Manoharetal,Lutyetal} can then be used to
estimate the relative strengths of the various couplings in the theory.  Since
the geometrical factors present in loop calculations determine the suppression
of quantum effects relative to classical effects the condition that all
interactions are strong at the fundamental scale is dependent on the number of
dimensions that the fields of the theory propagate in. NDA in the context of
extra dimensions was first studied in \cite{Chacko:1999hg}.  Here we are
interested in NDA in \emph{warped} extra dimensions. We will be interested in
cases where the cutoff $\Lambda$ is larger than the curvature scale $k$. This
means that the parameters of the theory are determined by physics at much
shorter distances than the curvature scale.  Therefore the only effect of the
warping is that the local cutoff $\Lambda e^{-k y }$ will now vary as a
function of location in the fifth dimension.  These assumptions of NDA lead to
a Lagrangian of the form below.  \be
\label{eqn:ndaform}
S\sim\int d^5 x\sqrt{-g} \left(\frac{\Lambda^5}{\epsilon\,
    l_5}\hat{\mathcal{L}}_{\mathrm{bulk}}(\hat{\Phi},\frac{\partial}{\Lambda})
  + \delta(y-y_{\mathrm{brane}}) \frac{\Lambda^4}{\epsilon\, l_4}
  \hat{\mathcal{L}}_{\mathrm{brane}}(\hat{\Phi},\hat{\phi},
  \frac{\partial}{\Lambda})\right).
\ee
Here $l_D=2^D\pi^{D/2}\Gamma(D/2)$ is the loop factor in $D$ dimensions.  The
parameter $\epsilon$ {\cite{Chacko:1999mi}}
is a measure of the amplitudes of one
loop processes relative to tree level processes, or more generally (N+1)
loop processes relative to N loop processes. $\epsilon$ lies in the range
$0\leq\epsilon\leq 1$ with $1$ being the strong coupling limit and $0$
being the free theory.  Hatted fields are dimensionless and all the
couplings in $\hat{\mathcal{L}}_{\mathrm{bulk}}$ and
$\hat{\mathcal{L}}_{\mathrm{brane}}$ are order one at the high scale.

We now wish to determine the sizes of various parameters in the theory.  
In going from the 5 dimensional action to the effective 4 dimensional
theory it is necessary to take into account the $y$ profile of the bulk
fields.  One simplification that occurs in the case of bulk gauge fields
is that due to 4 dimensional gauge invariance the gauge boson zero-mode is
flat in the extra dimension.  SUSY is unbroken at the compactification
scale so the gaugino acquires a profile such that the whole gauge action
is $y$ independent.  Therefore gauge and gaugino zero modes behave like
flat space zero modes. 

Let us first relate the strong coupling scale to the 5 dimensional Planck
scale.  Recall the Einstein-Hilbert action in 5 dimensions,
\be
\int d^5 x\sqrt{-g}M_5^3 R,
\ee
where $R$ contains 2 derivatives of the metric.  Comparing this to
Eqn.~(\ref{eqn:ndaform}) we see that 
\be
\frac{\Lambda^3}{\epsilon\, l_5}=M_5^3.
\ee

A similar analysis can be carried out to relate the bulk gauge coupling to the
effective 4 dimensional gauge coupling.  The zero mode gauge field action is
(schematically) given by,
\be
S\sim \int d^4 xdy \frac{\Lambda^5}{\epsilon\, l_5}
\left(\left(\frac{\partial}{\Lambda} \hat{A}^{(0)}\right)^2 +
  \frac{\partial}{\Lambda}\hat{A}^{(0)3}+\hat{A}^{(0)4}\right), 
\ee
due to gauge invariance there is no dependence on the $y$ coordinate.
In canonical normalization the 4 dimensional gauge fields are given by,
\be
A_\mu\sim \left(\frac{\Lambda^3 2\pi R}{\epsilon
    l_5}\right)^{1/2}\hat{A}^{(0)}_\mu, 
\ee
and so the 4 dimensional gauge coupling is given by,
\be\label{eqn:fourdgaugecoupling}
g_4^2=\frac{\epsilon\, l_5}{2\pi \Lambda R}. 
\ee
We demand that the 4D gauge coupling is order one and that the extra dimension
is large enough that the infrared scale is order $1$ TeV, i.e. $k\pi R\sim
30$.  Using these relations along with Eqn.(\ref{4DPlanckscale}) gives, 
\be
M_4^2\sim \frac{\Lambda^2}{30}.  
\ee 
It will be useful to know the ratio
$k/\Lambda$ which is, 
\be 
\frac{k}{\Lambda}\sim\frac{30}{\epsilon\, l_5}\sim
\frac{1}{25\epsilon}.  
\ee 
By definition $\epsilon<1$, in addition we require
$k/\Lambda<1/4$ placing lower limits on $\epsilon$.  Thus, $\epsilon$ lies in
the range $1/6<\epsilon<1$ and $1/25<k/\Lambda<1/4$.  Now we discuss the
relative sizes of the various SUSY breaking effects with the NDA assumption.

First consider the supersoft operator that generates the Dirac gaugino
masses.  The relevant piece of the action is given by,
\bea
S& \sim& \int d^5 x\sqrt{-g}\left[ \frac{\Lambda^5}{\epsilon\, l_5} \int
\frac{d^2\theta}{\Lambda} \hat{W}_\alpha \hat{W}^\alpha + \delta (y-\pi R)
\frac{\Lambda^4}{\epsilon\, l_4}\left( \int \frac{d^2\theta}{\Lambda}
\hat{W}^\prime_\alpha \hat{W}^{\prime\alpha}\right.\right.\\ &&\left.\left.+
\int \frac{d^2\theta}{\Lambda} \hat{W}_\alpha^\prime \hat{W}^\alpha \hat{A}
+ \int \frac{d^4\theta}{\Lambda^2} \hat{A}^\dagger
\hat{A}\right)\right] \\
&\sim& \int d^4 x\frac{2\pi R \Lambda^5}{\epsilon
  l_5}\hat{\lambda}\frac{\partial}{\Lambda} \hat{\lambda} +
\frac{\Lambda^4}{\epsilon\, l_4} e^{-4\sigma(\pi R)} \left(e^{3\sigma(\pi R)/2}
  \hat{D}^\prime \hat{\lambda} \hat{\psi}_A + \hat{D}^{\prime 2}\right.\\
&&\left.+  e^{\sigma(\pi R)} \hat{\bar{\psi}}_A\frac{\partial}{\Lambda}
  \hat{\psi}_A \right)+\ldots, 
\eea
where in the second line we have inserted the $y$ profiles of the fields and
integrated over the extra dimension.  Rescaling fields so that they are
canonically normalized and using the fact that the four dimensional gauge
coupling is order 1, Eqn.(\ref{eqn:fourdgaugecoupling}), the Dirac mass term
is, 
\be\label{eqn:appendixdiracmass}
M_D\sim e^{\sigma(\pi R)}\frac{\overline{D}}{\Lambda}\lambda \psi_A.
\ee
Note that $\overline{D}=D^\prime e^{-2\sigma(\pi R)}$.  Here $\overline{D}$
should be thought of as the VEV in warped down units and $D^\prime$ is the VEV
in the 5D theory where all mass scales are defined at the 5D Planck scale.
This Dirac gaugino mass in turn gives supersoft contributions to the scalar
masses at one loop,
\be
m_{SS}^2\sim\frac{1}{l_4}
\left(\frac{\overline{D}}{\Lambda}\right)^2 e^{2\sigma(\pi R)}.
\ee

The non-zero $D$-term VEV is generated from a VEV for the fields $P$ and
$\bar{P}$ charged under the $U(1)^\prime$.  We can relate the sizes of the
$\overline{D}$ and $\langle P \rangle$ by considering the operator $\hat{P}^\dagger
e^{\hat{V}^\prime} \hat{P}$ and rescaling $\hat{P}$ to give it canonical
kinetic term to find, 
\be
\langle \overline{D} \rangle=\sqrt{\epsilon\, l_4}\langle P\rangle^2.
\ee

Hard SUSY breaking operators of the form ({\ref{hardbreaking}}) localised on
the UV brane can alter the relative couplings of gauginos and gauge bosons to
matter.  These in turn generate flavour diagonal corrections to the scalar
masses at one loop. We now estimate the size of these corrections using NDA.
The operator we consider is, 
\be \int d^5 x \sqrt{-g} \delta(y-\pi R)
\frac{\Lambda^4}{\epsilon\, l_4} \int \frac{d^4\theta}{\Lambda^2}
\hat{P}^\dagger e^{\hat{V}^\prime} \hat{P} \hat{W}_\alpha
\frac{D^2}{\Lambda}\hat{W}^\alpha.  
\ee 
In terms of canonically normalized fields the correction to the gaugino
kinetic term is, 
\be 
\frac{\overline{D}^2}{\Lambda^4}e^{4\sigma (\pi R)}\lambda \partial \lambda.  
\ee 
This is a correction to the gauge coupling, $1/g^2\rightarrow
1/g_0^2+(\overline{D}e^{{2\sigma (\pi R)}}/\Lambda^2)^2$.  This leads to a one
loop quadratically divergent contribution to the scalar mass which is cutoff
at the mass of the lightest KK state. So, 
\be 
m_{HB}^2=\frac{1}{l_4} \left(\frac{\overline{D}}{\Lambda^2}\right)^2 
e^{4\sigma(\pi R)} g_4^4 \left(ke^{-\sigma(\pi R)}\right)^2 =
\frac{1}{l_4}\left(\frac{\overline{D}}{\Lambda}\right)^2
\left(\frac{k}{\Lambda}\right)^2 e^{2\sigma(\pi R)}, 
\ee 
which is smaller than the supersoft contribution by a
factor of $\left(k/\Lambda\right)^2\sim \left(1/25\epsilon\right)^2$.

There is a two loop logarithmically divergent correction to the soft scalar
masses arising from the soft mass for the adjoint Eqn.({\ref{Asoftmass}}), as
shown in ~\cite{Poppitz:1997xw}. The logarithm is cut-off close to the mass of
the lightest KK state.  This correction will be negative provided the
logarithm is larger than order unity.  
The operator that leads to the adjoint soft mass has the form
\be\label{eqn:ptgenerator}
\int d^5 x \sqrt{-g} \delta (y-\pi R) \int \frac{d^4\theta}{\Lambda^2}
\hat{P}^\dagger e^{\hat{V}^\prime}\hat{P} \hat{A}^\dagger \hat{A} 
\rightarrow \epsilon\, l_4 e^{2\sigma (\pi R)}\frac{\overline{D}^2}{\Lambda^2}
A^2 \ee
The leading contribution to the scalar mass is~\cite{Poppitz:1997xw},
\be
m_{\tilde{q}}^2\sim -\frac{1}{l_4}\epsilon \left(\frac{\overline{D}}{\Lambda}
\right)^2 e^{2\sigma (\pi R)}\log\frac{ke^{-\sigma(\pi R)}}{M_D}
\ee
%
%As long as the log is smaller than $\mathcal{O}(6)$ we can suppress this
%contribution relative to the supersoft contribution, we define their ratio to
%be $\eta<1$.

Finally we wish to know the size of the other possible supersoft operator,
$W^\prime_\alpha W^{\prime\alpha} A^2$, this operator splits the mass squared
of the real and imaginary parts of the adjoint scalar, pushing one positive
and one negative. Using NDA the size can be estimated,
\be
\int d^5 x\sqrt{-g}\delta(y-\pi R)\int \frac{d^2\theta}{\Lambda^2}
\frac{\hat{W}^\prime_\alpha \hat{W}^{\prime\alpha}}{\Lambda^2}\hat{A}^2.
\rightarrow \epsilon l_4\left(\frac{\overline{D}}{\Lambda}\right)^2 e^{2\sigma(\pi R)}A^2
\ee
Notice that this is the same size as the positive contribution from
Eqn.(\ref{eqn:ptgenerator}), so that both the real and imaginary component 
of each
scalar adjoint can naturally have a net positive mass squared.

\end{document}